\documentclass[12pt]{article}

\usepackage{graphicx}

\newcommand{\be}{\begin{equation}}
\newcommand{\ee}{\end{equation}}
\newcommand{\bq}{\begin{eqnarray}}
\newcommand{\eq}{\end{eqnarray}}
 
\newcommand{\D}{\mathrm{d}}
\newcommand{\E}{\mathrm{e}}

\def\lsim{\mathrel{\rlap{\lower4pt\hbox{\hskip1pt$\sim$}}\raise1pt\hbox{$<$}}}
\def\gsim{\mathrel{\rlap{\lower4pt\hbox{\hskip1pt$\sim$}}\raise1pt\hbox{$>$}}}
\def\Vec#1{\mathpalette{\VVec}{#1}}                  
\def\VVec#1#2{\mbox{\boldmath$#1#2$\unboldmath}}

\newcommand{\Strut}{\rule[-1.7ex]{0pt}{4.7ex}}        
\def\anti#1{\mathpalette{\@anti}{#1}#1}
\def\@anti#1#2{\sbox0{$#1#2$}
  \makebox[0pt][l]{$#1\kern.30\ht0\overline{\kern-.35\ht0\phantom{#2}}$}}

\begin{document}
\null 

\vspace{1cm} 
\begin{center}
{\Large \bf Phenomenology of SIDIS unpolarized \\  cross sections  and azimuthal 
asymmetries}\footnote{Invited review talk at ``Transversity 2011'', Veli Lo{\v s}inj, Croatia.}
\end{center} 

\vspace{1cm}

\begin{center}
{\large Vincenzo Barone} 

\vspace{0.5cm} 
{\it Di.S.I.T., Universit{\`a} del Piemonte 
Orientale ``A. Avogadro''; \\
INFN, Gruppo Collegato di Alessandria,  15121 Alessandria, Italy}

\end{center}

\vspace{2cm} 

\begin{center} 
 {\bf Abstract}
\end{center}

\noindent{\small 
I review the phenomenology of unpolarized cross 
sections and azimuthal asymmetries in semi-inclusive 
deeply inelastic scattering (SIDIS).  The general theoretical 
framework is presented and the validity of the Gaussian model 
is discussed. A brief account of the existing analyses is provided.  
}

\section{Introduction}

It has been known since the early years of quantum chromodynamics
 that azimuthal asymmetries 
in unpolarized processes are generated by gluon radiation and splitting.   
The observation of these asymmetries was in fact proposed as a test of perturbative 
QCD~\cite{Georgi:1977tv}. The first experimental results date back to the Eighties and showed 
that unpolarized semi-inclusive deeply inelastic scattering (SIDIS)
indeed displays non vanishing azimuthal modulations of the type $\cos \phi_h$ and $\cos 2 \phi_h$   
 \cite{Aubert:1983cz,Arneodo:1986cf}. 

While the pQCD processes account for these asimmetries at large transverse momenta, in the  
low--$P_{h \perp}$ region it is the intrinsic transverse motion 
of quarks that plays a key r{\^o}le (for a recent review see \cite{Barone:2010zz}). 
There are two non-perturbative sources of 
azimuthal asymmetries in unpolarized processes: 1) the Cahn effect \cite{Cahn:1978se,Cahn:1989yf}, 
a purely kinematic correction due to the quark transverse momentum: 2) 
the Boer-Mulders effect \cite{Boer:1997nt}, arising from the correlation 
between the transverse momentum and the transverse spin of quarks inside an 
unpolarized nucleon.

\section{Theoretical framework}

We will consider unpolarized SIDIS, 
$l(\ell) + N(P) \to l(\ell') + h(P_h) + X(P_X)$, in the current 
fragmentation region. At low $P_{h \perp}$ the cross section of this process factorizes 
into a transverse-momentum dependent quark distribution (TMD) and a fragmentation 
function \cite{Ji:2004xq}. 

\begin{figure}[t]
\vspace{-1.5cm}
\centering
\includegraphics[width=0.35\textwidth,angle=-90]
{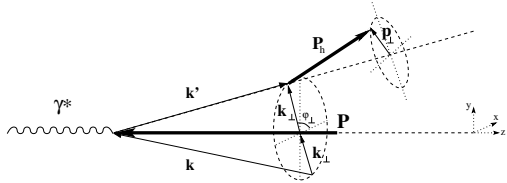}
\caption{Transverse kinematics in the photon-nucleon collinear frame.}
\label{tr_kin}
\end{figure}

Let us consider the transverse kinematics in a reference frame where 
the virtual photon and the nucleon are collinear (we call it $\gamma^* N$ 
collinear frame, see fig.~\ref{tr_kin}). 
The transverse momenta are denoted as follows: 
$\Vec k_{\perp}$,  transverse momentum of the initial quark; 
$\Vec p_{\perp}$,  transverse momentum of the hadron w.r.t. to the fragmenting quark;
$\Vec P_{h \perp}$,  transverse momentum of the hadron w.r.t. to the $\gamma^*N$ axis. 
Neglecting $1/Q^2$ corrections, these momenta are related by 
\be
 \Vec p_{\perp} \simeq \Vec P_{h \perp} - z_h \, \Vec k_{\perp}.
\label{mom_rel} 
\ee

The SIDIS structure functions are actually expressed in terms of the 
quark transverse momentum $\Vec k_T$ defined in the frame where the produced hadron $h$ and the 
nucleon are collinear ($hN$ collinear frame). However, if one neglects 
$1/Q^2$ corrections, the transverse momenta in the $\gamma^*N$ and  
$hN$ collinear frame coincide: $\Vec k_{\perp} \simeq \Vec k_T$.

The general expression of the cross section for unpolarized 
SIDIS is \cite{Mulders:1995dh,Bacchetta:2006tn}
 \bq
& & 
 \frac{\D \sigma}{\D x_B  \, \D y  \, \D z_h\,    
\D \phi_h \,  \D  \Vec P_{h \perp}^2 } = 
 \frac{2 \pi \, \alpha_{\rm em}^2}{ x_B y Q^2}
\, 
 \left \{ 
 (1 - y + \frac{1}{2} y^2) \,  F_{UU, T} 
+ (1 - y) \,  F_{UU, L} \right.
\nonumber \\
& & \hspace{0.5cm} + \left. 
 (2 - y) \sqrt{1 - y} \, \cos \phi_h 
 F_{UU}^{\cos \phi_h}
+  (1 - y) \, \cos 2 \phi_h \,  F_{UU}^{\cos 2 \phi_h} \Strut 
\right \}.
\label{cross}
\eq
The azimuthal asymmetries are defined as: 
\be
A^{\cos \phi_h} = 2 \, 
\langle \cos \phi_h \rangle =
2 \, \frac{\int \D \phi_h  \cos \phi_h \, 
\D \sigma }{\int \D \phi_h \,  \D \sigma }, 
\ee
and similarly for $A^{\cos 2 \phi_h}$. 

In the extended parton model, the structure functions appearing 
in eq.~(\ref{cross}) are given by convolutions of transverse momentum 
dependent quark distributions and fragmentation functions. Up to order $1/Q$ one has
\bq
& & 
F_{UU,T} =  {\mathcal C} \, \left [f_1 D_1 \right ],
\label{fuut} \\
& & 
F_{UU, {\rm Cahn}}^{\cos \phi_h} = \frac{2 M}{Q}   
\, {\mathcal C} \left [ 
-   \frac{\hat{\Vec h} \cdot \Vec k_T}{M} \, f_1 D_1 \right ],  
\label{fuucahn} \\ 
& &  
F_{UU, {\rm BM}}^{\cos \phi_h} = \frac{2 M}{Q}   
\, {\mathcal C} \left [ - \frac{(\hat{\Vec h} \cdot \Vec k_T') 
\,  \Vec k_T^2}{M_h M^2} \, h_1^{\perp} \, H_1^{\perp} \right ],
\label{fuubm} \\
& & 
F_{UU, {\rm BM}}^{\cos 2 \phi_h} = 
{\mathcal C} 
\left [ -\frac{2 (\hat{\Vec h} \cdot \Vec k_T) (\hat{\Vec h} 
\cdot \Vec k_T') - \Vec k_T \cdot \Vec k_T'}{M M_h} 
\, h_1^{\perp} \, H_1^{\perp} \right ],  
\label{fuubm2}
\eq 
where 
${\mathcal C} \, [w \, f \, D]$ 
is a transverse momentum convolution. 

Equation~(\ref{fuucahn}) represents the Cahn  contribution to the 
$\cos \phi_h$ structure function. It originates from the 
elementary lepton-quark cross section in presence of 
transverse momenta, which reads
\be
\D \hat{\sigma} \sim  \frac{1}{y^2} \, 
\left [ \left (1 - 4 \frac{{k}_{\perp}}{Q} \sqrt{1 - y} \, \cos \varphi \right ) + 
\left (1 - 4 \frac{{k}_{\perp}}{Q} \, \frac{\cos \varphi}{\sqrt{1 - y}} \right )  
\right ]  + {\mathcal O}\left (\frac{{k}_{\perp}^2}{Q^2} \right ). 
\ee

Equations~(\ref{fuubm}) and (\ref{fuubm2}) are the Boer-Mulders (BM) contributions 
to the azimuthal modulations of unpolarized SIDIS. They couple the Boer-Mulders 
function $h_1^{\perp}$ to the Collins function $H_1^{\perp}$ \cite{Collins:1992kk}, which 
describes  the fragmentation of transversely polarized quarks into an unpolarized 
hadron. Notice that the BM effect contributes to $\cos \phi_h$ at order 
$1/Q$ ({\it i.e.},  at twist three) and to $\cos 2 \phi_h$ at leading twist.

At twist three one should also take into account 
quark-gluon interactions, which give rise to the so-called ``tilde'' distributions.   
Thus $F_{UU}^{\cos \phi_h}$ gets the additional term \cite{Bacchetta:2006tn}
\bq
F_{UU}^{\cos \phi_h} &=& \frac{2 M}{Q}   
\, {\mathcal C} \left [ - \frac{\hat{\Vec h} \cdot \Vec k_T'}{M_h} 
\left ( x \tilde{h} H_1^{\perp} + \frac{M_h}{M} 
\ f_1 \frac{\tilde{D}^{\perp}}{z} \right ) 
\right. 
\nonumber \\
& & 
- \left. \frac{\hat{\Vec h} \cdot \Vec k_T}{M} 
\left ( x \tilde{f}^{\perp} D_1 + \frac{M_h}{M} 
h_1^{\perp} \frac{\tilde{H}}{z} \right ) \right ]. 
\label{fuu_tilde}
\eq 
We recall however that 
TMD factorization is not proven beyond leading twist, so eq.~(\ref{fuu_tilde}) 
must be taken with a grain of salt.

The Cahn effect produces at order $1/Q^2$ (twist four) 
a further contribution to the $\cos 2 \phi_h$ structure function,  
\be
F_{UU, {\rm Cahn}}^{\cos 2 \phi_h} = 
\frac{M^2}{Q^2} \, 
{\mathcal C} 
\left [ \frac{2 (\hat{\Vec h} \cdot \Vec k_T)^2 -  \Vec k_T^2}{M^2} 
 f_1 \, D_1 \right ]. 
\ee
This expression incorporates only part of the $1/Q^2$ kinematic corrections. 
Besides them, there are also
dynamical $1/Q^2$ corrections arising from twist-four TMD's and 
fragmentation functions. 
Therefore,  $F_{UU, {\rm Cahn}}^{\cos 2 \phi_h}$ is just 
an approximate
estimate of the full ${\mathcal O}(1/Q^2)$ structure function. 

In summary, ignoring unknown higher-twist terms, 
the two azimuthal asymmetries of unpolarized SIDIS are symbolically given by:   
\be
  \langle \cos \phi_h \rangle = \frac{1}{Q} \, \mbox{\rm Cahn} + 
\frac{1}{Q} \, \mbox{\rm BM} ,
\;\;\;\;\;\;
 \langle \cos 2 \phi_h \rangle = \mbox{\rm BM} + 
\frac{1}{Q^2} \, \mbox{\rm Cahn}.   
\label{asym_symb}
\ee

\section{The Gaussian approach}

In most phenomenological analyses, 
the TMD's and fragmentation functions 
are parametrized by Gaussians: 
\be
 f_1 (x, \Vec k_{\perp}^2) = {\mathcal A}\, f_1 (x) \, 
\E^{- \Vec k_{\perp}^2/\overline{\Vec k_{\perp}^2}} 
\;\;\;\;
 D_1 (z, \Vec p_{\perp}^2) = {\mathcal B} \, D_1 (z) \, 
\E^{- \Vec p_{\perp}^2/\overline{\Vec p_{\perp}^2}}.
\label{gauss}
\ee
Here $x$ is the fraction of the nucleon light-cone momentum 
carried by the quark, and $z$ is the fraction of the light-cone momentum 
of the struck quark carried by the final hadron. These two variables 
coincide with $x_B$ and $z_h$, respectively, modulo $1/Q^2$ corrections. 
We stress that there is a strong assumption -- with no 
fundamental basis -- behind the Gaussian
model, namely that distributions factorize in $x$ and $k_{\perp}$.

In principle, one has to distinguish the Gaussian widths 
$\overline{\Vec k_{\perp}^2}$ and $\overline{\Vec p_{\perp}^2}$  
from the average squared momenta, defined as 
\be
\langle \Vec k_{\perp}^2 \rangle \equiv \int \D^2 \Vec k_{\perp} \, 
\Vec k_{\perp}^2  \, f_1 (x, \Vec k_{\perp}^2), 
\;\;\;\;
\langle \Vec p_{\perp}^2 \rangle \equiv \int \D^2 \Vec p_{\perp} \, 
\Vec p_{\perp}^2  \, D_1 (z, \Vec p_{\perp}^2). 
\label{average}
\ee
Average values coincide with the Gaussian widths,  
$\langle \Vec k_{\perp}^2 \rangle = \overline{\Vec k_{\perp}^2}, 
 \langle \Vec p_{\perp}^2 \rangle = \overline{\Vec p_{\perp}^2}$,  
only if we integrate over $k_{\perp}$ 
and $p_{\perp}$  between 0 and $\infty$.

Bounds on $x \equiv k^-/P^-$ and $k_{\perp}$ can be obtained in the quark parton model
by inserting a set of physical intermediate states in the quark-nucleon amplitudes \cite{Jaffe:1983}. 
Since the momentum of these states is 
\be
P_n^{\mu} = \left (\frac{x M^2 - \Vec k_{\perp}^2}{2 x P^-}, (1 - x) P^-, - \Vec k_{\perp} \right ),
\ee 
the condition $M_n^2 \geq 0$ yields
\be
\Vec k_{\perp}^2 \leq x (1 - x) M^2.  
\label{bound_transv}
\ee
This is the bound on the truly {\it intrinsic} quark transverse momentum
\cite{Sheiman:1980}
(for recent discussions see \cite{Dalesio:2010,Zavada:2011}). 
Numerically the upper limit  (\ref{bound_transv}) 
is $\Vec k_{\perp}^2 < 0.25$ GeV$^2$. Clearly, 
the average squared momentum $\langle \Vec k_{\perp}^2 \rangle$ must be smaller than this 
value. On the other hand, various phenomenological analyses give 
much larger figures, 
$\langle \Vec k_{\perp}^2 \rangle \sim 0.25 - 0.40$ GeV$^2$.

Is there a contradiction between these results?
The answer is that 
the bound (\ref{bound_transv}) actually refers to a ``static'' nucleon ($Q^2 =0$), 
whereas the value of  $\langle \Vec k_{\perp}^2 \rangle$ extracted from experiments 
effectively accounts for the ``non-intrinsic'' transverse momentum,  radiatively generated at a given 
$Q^2$. As a consequence, 
 $\langle \Vec k_{\perp}^2 \rangle$ is not a fixed, universal, quantity.

Consider now the unpolarized and angle-independent structure function: 
\be
 F_{UU} = \sum_a e_a^2 \int \D^2 \Vec k_{\perp} 
\int \D^2 \Vec p_{\perp} \, \delta^2 (\Vec p_{\perp} + z_h \Vec k_{\perp} - \Vec P_{h \perp}) 
\, f_1^a (x_B, \Vec k_{\perp}^2) \, D_1^a (z_h, \Vec p_{\perp}^2) 
\ee
Using Gaussian parametrizations and integrating between 0 and $\infty$, one gets 
\be
F_{UU} = \sum_a e_a^2 \, f_1^a(x_B)\, D_1^a (z_h) \, 
\frac{\E^{-\Vec P_{h \perp}^2/\overline{\Vec P_{h \perp}^2}}}{\pi \, 
\overline{\Vec P_{h \perp}^2}} 
\ee
with 
\be
\overline{\Vec P_{h \perp}^2} = \overline{\Vec p_{\perp}^2} + z_h^2 \, 
\overline{\Vec k_{\perp}^2} 
\label{pop_rel}
\ee
This is a popular relation appearing in many analyses (where it is 
often quoted as a relation between average momenta). It must used with 
some caution. In fact: {\it i)} it is 
invalidated by a possible cutoff on $\Vec k_{\perp}^2$; 
{\it ii)} it has (complicated) $1/Q^2$ corrections.  {\it iii)}  
due to experimental cuts,  $\overline{\Vec P_{h \perp}^2}$ differs 
from the measured $\langle \Vec P_{h \perp}^2 \rangle$.

\begin{figure}[t]
\centering
\includegraphics[width=0.35\textwidth,angle=0]
{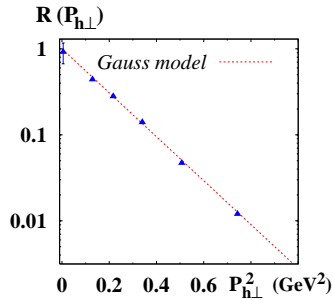}
\caption{The ratio $R(P_{h \perp}) = \D \sigma_{UU}(P_{h \perp})/\D \sigma_{UU} (0)$: 
Gaussian model vs. CLAS data (from \cite{Schweitzer:2010tt}).}  
\label{fig_metz1}
\end{figure}

\section{Transverse-momentum dependence of cross sections}

The Gaussian Ansatz for the unpolarized SIDIS cross section, 
\be
\frac{\D \sigma_{UU} ({\rm P}_{h \perp})}{\D z \, \D \Vec P_{h \perp}^2} = 
 \frac{\D \sigma_{UU} (0)}{\D z \, \D \Vec P_{h \perp}^2}
\, \E^{- \Vec P_{h \perp}^2/\overline{\Vec P_{h \perp}^2}}, 
\ee
has been tested by Schweitzer, Teckentrup and Metz \cite{Schweitzer:2010tt} 
in the CLAS kinematics ($x_B = 0.24$, $z_h = 0.34$, $Q^2 = 2.4$ GeV$^2$) \cite{Osipenko:2008rv}.  
As shown in fig.~\ref{fig_metz1}, the description of data is 
very good, with a Gaussian width  
$\overline{\Vec P_{h \perp}^2} = 0.17$ GeV$^2$. 

\begin{figure}[tb]
\vspace{0.5cm}
\centering
\includegraphics[width=0.40\textwidth]{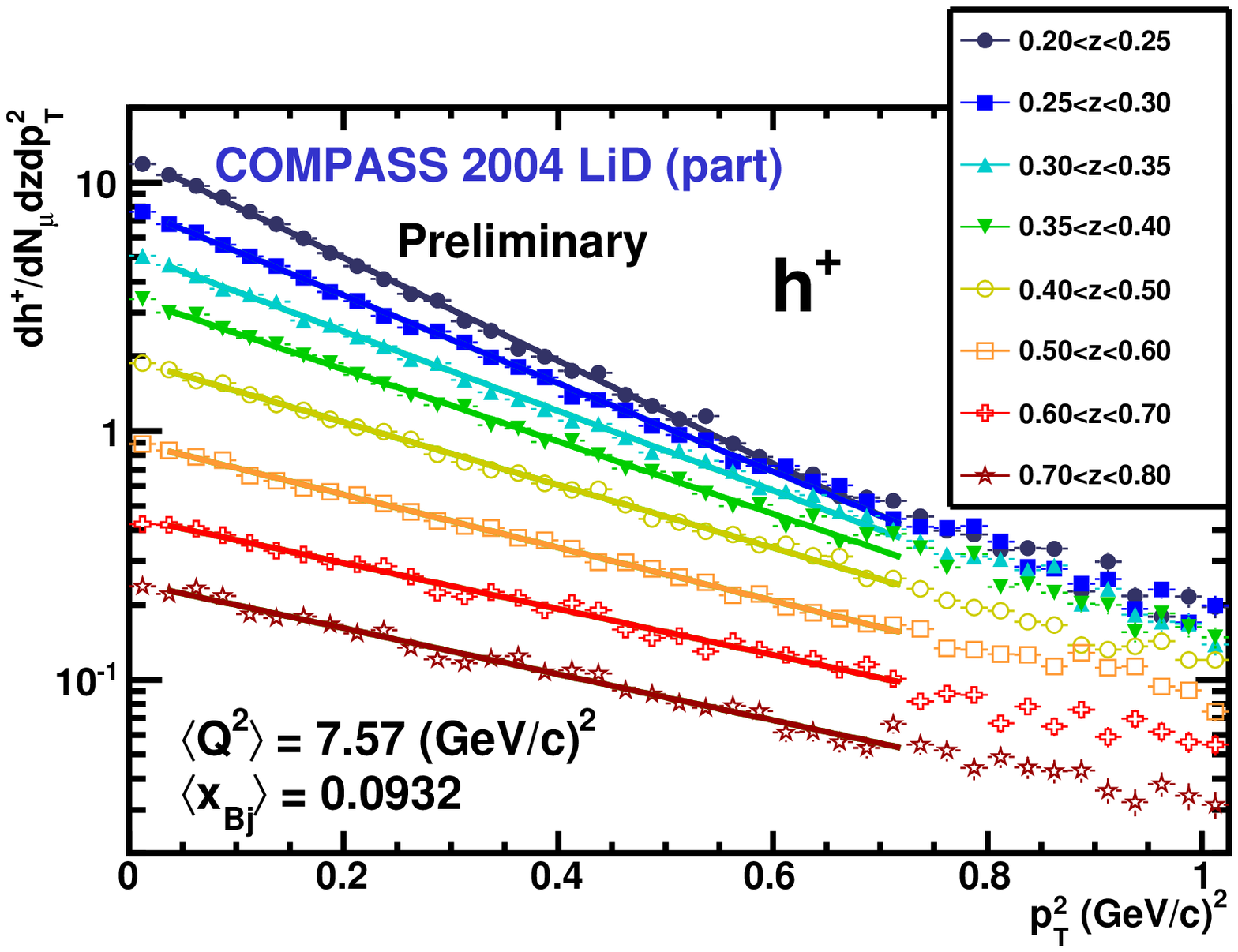} 
\hspace{0.5cm}
\includegraphics[width=0.40\textwidth,angle=0]
{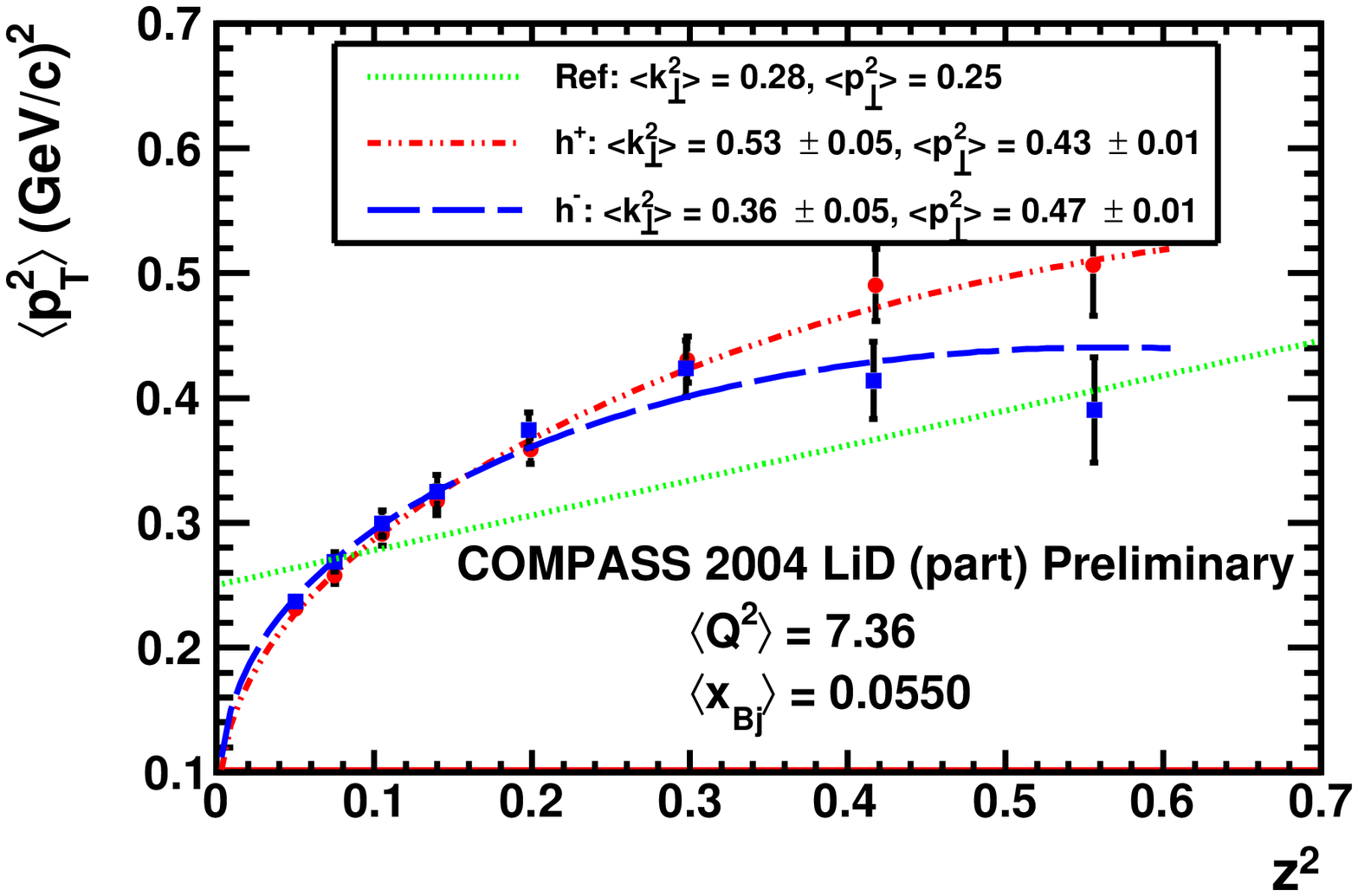}
\caption{Left: $P_{h \perp}^2$ distributions fitted by Gaussians (COMPASS data) 
\cite{Rajotte:2010ir}. Right: Fitted $\overline{P_{h \perp}^2}$ (called 
$\langle p_T^2 \rangle$ in the figure) vs. $z_h^2$ (COMPASS data) 
\cite{Rajotte:2010ir}.} 
\label{fig_compass1}
\end{figure}

The results of a Gaussian analysis of 
the $\Vec P_{h \perp}^2$ distributions measured by COMPASS  
are displayed in fig.~\ref{fig_compass1} (left) \cite{Rajotte:2010ir}. 
The fitted $\overline{\Vec P_{h \perp}^2}$ is found to have     
a mild dependence on $x_B$ and to increase with $Q^2$.  
A check of the relation (\ref{pop_rel}) has been also performed,   
with the result shown in fig.~\ref{fig_compass1} (right). 
One sees a clear 
departure from a linear dependence of 
$\overline{\Vec P_{h \perp}^2}$
on $z_h^2$.  
The dashed and dot-dashed curves in the figure are fits 
of the type 
\[
\overline{ \Vec P_{h \perp}^2}  = 
z_h^{0.5} (1 - z_h)^{1.5} \, \overline{\Vec p_{\perp}^2} + z_h^2 \, 
\overline{\Vec k_{\perp}^2}
\]

\begin{figure}[t]
\centering 
\includegraphics[width=0.35\textwidth,angle=0]
{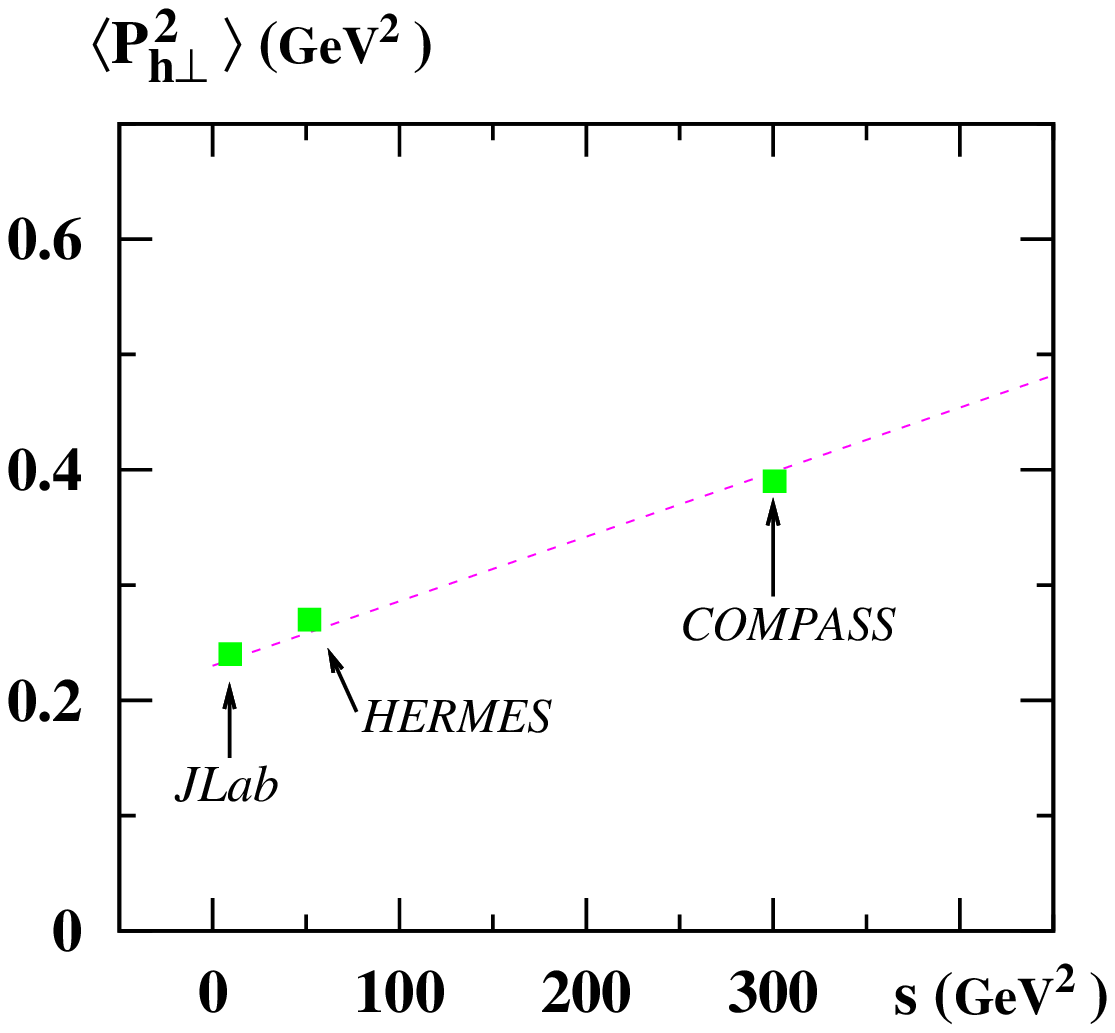} 
\hspace{0.5cm}
\includegraphics[width=0.42\textwidth,angle=0]
{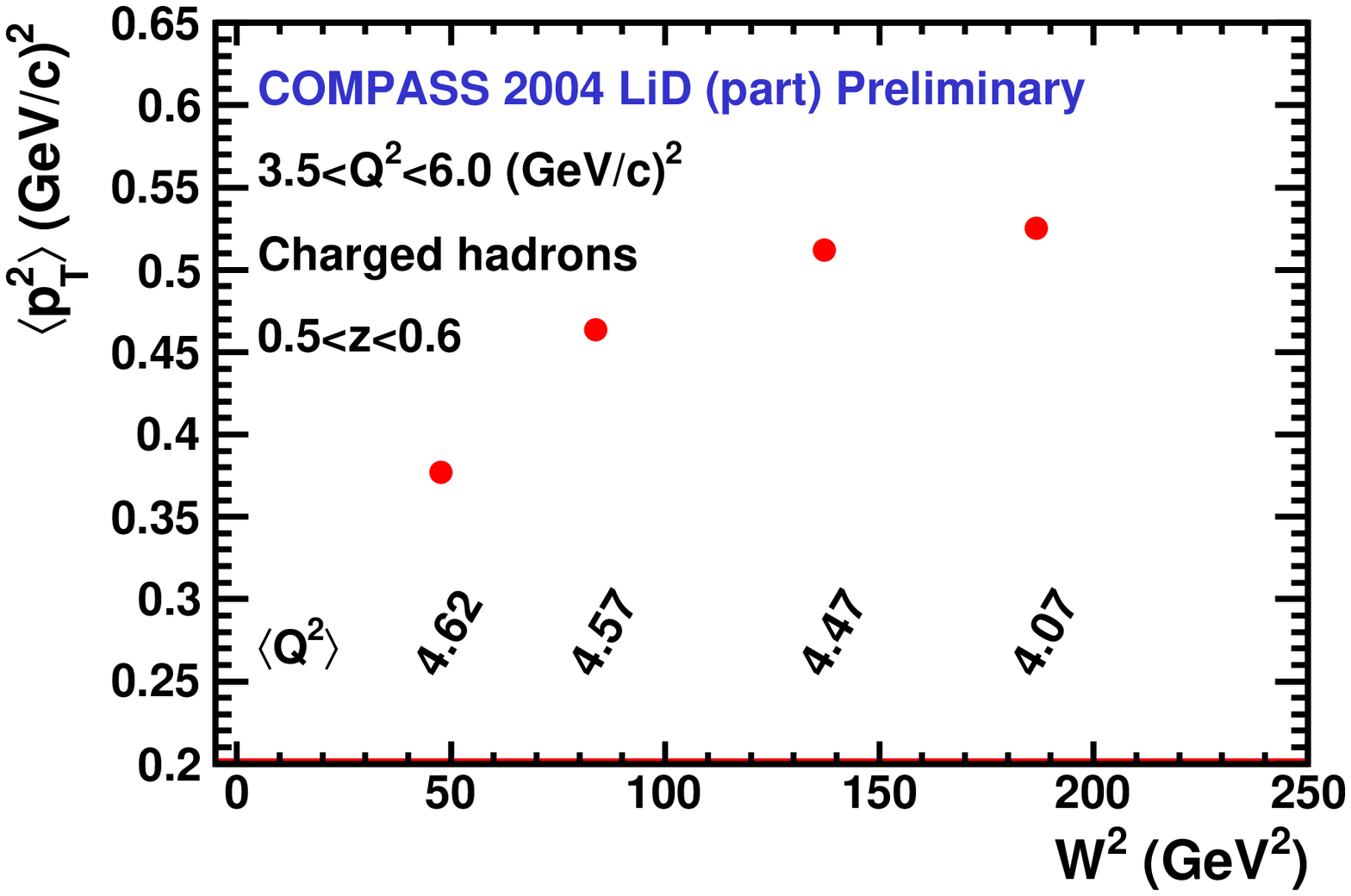}
\caption{Left: Dependence of $\langle \Vec P_{h \perp}^2 \rangle$ 
on the lepton-nucleon c.m. energy squared of various experiments 
(from \cite{Schweitzer:2010tt}). 
Right: Dependence of $\langle \Vec P_{h \perp}^2 \rangle$ 
on the photon-nucleon c.m. energy squared $W^2$ 
(COMPASS data, from \cite{Rajotte:2010ir}).}
\label{fig_metz2}
\end{figure}

Finally, it has been noticed in \cite{Schweitzer:2010tt} that the average 
transverse momentum of hadrons 
extracted in different experiments (JLab, HERMES, COMPASS)  
has an approximately linear rise in $s$ (fig.~\ref{fig_metz2}, left). This energy dependence 
is confirmed by a COMPASS 
analysis \cite{Rajotte:2010ir}, which shows that $\langle \Vec P_{h \perp}^2 \rangle$
increases with $W^2$ (fig.~\ref{fig_metz2}, right), which may be an indication of a possible 
broadening of $\langle \Vec k_{\perp}^2 \rangle$ 
and $\langle \Vec p_{\perp}^2 \rangle$. 
 
\section{Azimuthal asymmetries}

The first analysis of the $\cos \phi_h$ asymmetry in SIDIS 
in terms of the Cahn effect was presented in 2005 by Anselmino {\it et al.} 
\cite{Anselmino:2005nn}. They fitted the EMC data \cite{Aubert:1983cz}  
with a Gaussian model, obtaining the values 
$\overline{\Vec k_{\perp}^2} = 0.25$ GeV$^2$, 
$\overline{\Vec p_{\perp}^2} = 0.20$ GeV$^2$, but did not 
consider the Boer-Mulders term, which is of the same order 
as the Cahn contribution.  

The $\cos 2 \phi_h$ asymmetry was extensively studied 
in \cite{Barone:2008tn,Barone:2009hw}. Both the 
leading-twist Boer-Mulders contribution and the 
twist-4 Cahn term were included in the analysis. 
Signs and magnitudes of the Boer-Mulders function, 
extracted from a fit to HERMES \cite{Giordano:2009hi} and COMPASS \cite{Bressan:2009eu}  
preliminary data, were found to be in agreement with the 
theoretical expectations  (based on the impact-parameter approach, lattice studies  
and large $N_c$ arguments):
$h_{1}^{\perp u} \sim 2 f_{1T}^{\perp u}$, 
$h_{1}^{\perp d} \sim - f_{1T}^{\perp d}$. 

\begin{figure}[t]
\centering   
\includegraphics[width=0.40\textwidth,angle=-90]
{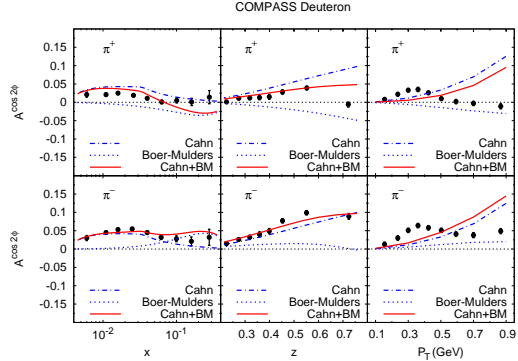} 
\caption{The fit of \cite{Barone:2009hw} to the COMPASS preliminary 
data on $\langle \cos 2 \phi_h \rangle$ with a deuteron target. }
\label{fig_compass_cos2phi}
\end{figure}

In particular, it was pointed out that, since the Cahn 
effect is the same for $\pi^+$ and $\pi^-$, and the 
favorite and unfavorite Collins functions 
are related by 
 $H_{1}^{\perp {\rm fav}} \approx - H_1^{\perp {\rm unf}}$, 
as shown by the fits to the Collins effect in SIDIS, 
a signature of the Boer-Mulders effect is   $\langle \cos 2 \phi_h \rangle_{\pi^-} > 
\langle \cos 2 \phi_h \rangle_{\pi^+}$. 
Another interesting output of the analysis in \cite{Barone:2008tn,Barone:2009hw} 
is that the Cahn contribution turns 
out to be relatively large in spite of being ${\mathcal O}(1/Q^2)$. 
An example of the fits presented in \cite{Barone:2009hw} is shown 
in fig~\ref{fig_compass_cos2phi}. 

The HERMES and COMPASS Collaborations have recently presented new 
preliminary data on azimuthal asymmetries in unpolarized SIDIS 
\cite{Giordano:2010zz,Sbrizzai:2010fi}. 
While the two measurements are in fairly good agreement for $\langle \cos \phi_h \rangle$, 
the COMPASS result for the 
$\cos 2 \phi_h$ asymmetry is systematically larger than 
the corresponding HERMES result.  

The CLAS Collaboration at JLab has released final results 
on the azimuthal modulations \cite{Osipenko:2008rv}. 
A large discrepancy is found between the data on 
$\langle \cos \phi_h \rangle$ and the Cahn effect predictions 
(extrapolated from other analyses). The $\langle \cos 2 \phi_h \rangle$ 
signal is non zero only in the small-$z_h$ region, where target fragmentation 
effects are important and may affect the interpretation of data.

From the theoretical viewpoint, the situation is quite unclear, as
only partial analyses are presently 
available. If we simply extrapolate them, we find that  
the Cahn contribution to $\langle \cos \phi_h \rangle$ is huge and largely overshoots 
the data, and that 
the Boer-Mulders contribution to $\langle \cos 2 \phi_h \rangle$ is small and  
apparently goes in the wrong direction, in the sense that it  
has an opposite sign in $\langle \cos \phi_h \rangle$ 
and $\langle \cos 2 \phi_h \rangle$, and therefore 
should produce a $\langle \cos \phi_h \rangle_{\pi^-}$ 
smaller than $\langle \cos \phi_h \rangle_{\pi^+}$, 
 whereas the data show a similar $\pi^+$-$\pi^-$ pattern for both 
asymmetries.

We have already said that the intrinsic transverse momentum 
of quarks is kinematically bounded. We expect 
that the radiatively generated transverse momentum 
should be limited as well. 
The effect of an upper bound on $k_{\perp}$ has been 
recently explored by Boglione, Melis and Prokudin \cite{Boglione:2011}. They derived a  
cutoff of the form $\Vec k_{\perp}^2 \leq \eta(x_B) Q^2$ in the $\gamma^* N$ center-of-mass frame. 
The consequence of setting this bound 
is that the Cahn contribution gets largely suppressed, while 
the BM contribution remains almost unaffected. This correction goes 
in the right direction, but further work is needed to establish a 
frame-independent condition on $k_{\perp}$.

\section{Conclusions}

There is nothing magic about the Gaussian approach. It is just a parametrization, 
which happens to work rather well for cross sections at low $P_{h \perp}$. 
Since this model lacks a solid basis, 
one should not be surprised 
if some simple regularities based on it turn out to fail.

The parameters $\overline{\Vec k_{\perp}^2}$ and 
$\overline{\Vec p_{\perp}^2}$ in the Gaussian distributions 
are likely to be $Q^2$ (and $W^2$) dependent. This means that 
they cannot be fixed once forever and used anywhere. 
One should do as for normal PDF's: take a TMD at small $Q^2$ from some 
model, evolve it and fit the data.

Although we have clear signals of relatively large azimuthal asymmetries, 
the situation is still unsettled. A combined state-of-the-art fit of $\langle \cos \phi_h \rangle$ 
and $\langle \cos 2 \phi_h \rangle$ would be highly desirable in order 
to understand the origin of these observables. 
We should  be  prepared to the possibility that the scheme 
(\ref{asym_symb}) 
might not work. We recall that the 
corrections to this scheme are of three types: 
1) ``genuine'' ({\it i.e.} ``tilde'') 
twist-3 contributions (originating from quark-gluon correlations); 
2) further $1/Q^2$ kinematic terms; 
3) dynamical twist-4 contributions.

A lesson we have learned is that 
higher twists are relevant in the presently explored 
region of SIDIS. Clearly, it would be important to disentangle 
them from leading twist contributions. In order to do so, 
a wider $Q^2$ range is needed -- an important task for  future experiments.  
In the meanwhile, the $1/Q^2$ contributions 
should be fully worked out, computing both the kinematic 
corrections and the target mass effects.

\vspace{1cm}

\begin{center}
{\bf Acknowledgments}
\end{center}

\noindent
I would like to thank Anna Martin and Franco Bradamante 
for their kind invitation to Transversity 2011, and 
Elena Boglione, Stefano Melis, Alexei Prokudin, Giulio Sbrizzai 
for useful discussions. 
This work is partially supported by the Italian Ministry of Education, 
University and Research (MIUR) in the framework of a Research Project 
of National Interest (PRIN 2008).

\end{document}